\documentclass[prc,preprint,amsmath,amssymb,showpacs]{revtex4}
\usepackage{verbatim}
\usepackage{bm}
\usepackage{epsfig}

\begin{document}

\title{Geometry of effective Hamiltonians}
\date{August 13, 2008}

\author{Simen Kvaal}
\email{simen.kvaal@cma.uio.no}
\affiliation{Centre of Mathematics for Applications, University of
  Oslo, N-0316 Oslo, Norway}

\begin{abstract}
  We give a complete geometrical description of the effective
  Hamiltonians common in nuclear shell model calculations. By
  recasting the theory in a manifestly geometric form, we reinterpret
  and clarify several points. Some of these results are hitherto
  unknown or unpublished. In particular, commuting observables and
  symmetries are discussed in detail. Simple and explicit proofs are
  given, and numerical algorithms are proposed, that improve and
  stabilize common methods used today.
\end{abstract}

\pacs{21.30.Fe,21.60.-n}

\maketitle

\newcommand{\CalH}{\mathcal{H}}
\newcommand{\CalV}{\mathcal{V}}
\newcommand{\CalW}{\mathcal{W}}
\newcommand{\CalM}{\mathcal{M}}
\newcommand{\CalP}{\mathcal{P}}
\newcommand{\CalQ}{\mathcal{Q}}
\newcommand{\CalE}{\mathcal{E}}
\newcommand{\CalF}{\mathcal{F}}
\newcommand{\CalG}{\mathcal{G}}
\newcommand{\CalA}{\mathcal{A}}
\newcommand{\CalB}{\mathcal{B}}
\newcommand{\CalX}{\mathcal{X}}
\newcommand{\CalY}{\mathcal{Y}}
\newcommand{\Real}{\operatorname{Re}}
\newcommand{\diag}{\operatorname{diag}}
\newcommand{\Span}{\operatorname{span}}
\newcommand{\Heff}{H_\text{eff}}
\newcommand{\eff}{{\text{eff}}}
\newcommand{\range}[2]{#1\negmedspace:\negmedspace#2}
\newcommand{\bra}[1]{\langle{#1}|}
\newcommand{\ket}[1]{|{#1}\rangle}
\newcommand{\braket}[1]{\langle{#1}\rangle}

\newcommand{\doubleunder}[2]{{\begin{smallmatrix}{#1}\\{#2}\end{smallmatrix}}}

\section{Introduction}

Effective Hamiltonians and interactions are routinely used in
shell-model calculations of nuclear spectra
\cite{Dean2004,Caurier2005,Navratil2000}. The published mathematical
theory of the effective Hamiltonian is complicated and usually focuses
on perturbation theoretical aspects, diagram expansions,
etc.\ \cite{Bloch1958,Brandow1967,Klein1974,Shavitt1980,Lindgren1974,Ellis1977,Lindgren1982,Suzuki1982,Kuo1981,Andreozzi1996}. 
In this article, we recast and reinterpret the basic elements of the
theory geometrically. We focus on the geometric relationship between
the exact eigenvectors $\ket{\psi_k}$ and the effective eigenvectors
$\ket{\psi_k^\text{eff}}$, both for the usual non-Hermitian
Bloch-Brandow effective Hamiltonian
\cite{Bloch1958,Brandow1967,Ellis1977,Dean2004}, and for the Hermitian effective
Hamiltonian \cite{VanVleck1929,Kemble1937,Klein1974,Suzuki1982}, which
we dub the canonical effective Hamiltonian due to its geometric
significance. This results in a clear geometric understanding of the
de-coupling operator $\omega = Q\omega P$, and a simple proof and
characterization of the Hermitian effective Hamiltonian in terms of
subspace rotations, in the same way as the non-Hermitian Hamiltonian is
characterized by subspace projections.

As a by-product, we obtain a simple and stable numerical algorithm to
compute the exact effective Hamiltonian.

The goal of effective interaction theory is to devise a Hamiltonian
$\Heff$ in a model space $\CalP$ of (much) smaller dimension $m$ than
the dimension $n$ of Hilbert space $\CalH$, with $m$ \emph{exact}
eigenvalues of the original Hamiltonian $H = H_0 + H_1$, where $H_1$ is
usually considered as a perturbation. 
The model space $\CalP$ is usually taken as the span of a few
eigenvectors $\{\ket{e_k}\}_{k=1}^m$ of $H_0$, i.e., the unperturbed Hamiltonian in a
perturbational view.

Effective Hamiltonians in $A$-body systems must invariably be
approximated (otherwise there would be no need for $\Heff$), usually
by perturbation theory, but a sub-cluster approximation is also
possible \cite{Navratil2000,Klein1974}. In that case, the exact
$a$-body canonical effective Hamiltonian is computed, where $a<A$.
From this, one extracts an effective $a$-body interaction and apply it to the
$A$-body system. In this case, we present a new algorithm for
computing the exact effective interaction that is conceptually and
computationally simpler than the usual one which relies on both matrix
inversion and square root \cite{Navratil2000,Suzuki1982}, as the only non-trivial
matrix operation is the singular value decomposition (SVD).

The article is organized as follows. In Sec.~\ref{sec:tools} we
introduce some notation and define the singular value decomposition of
linear operators and the principal angles and vectors between two
linear spaces. In Sec.~\ref{sec:effham} we define and analyze the
Bloch-Brandow and canonical effective Hamiltonians. The main part
consists of a geometric analysis of the exact eigenvectors, and forms
the basis for the analysis of the effective Hamiltonians. We also
discuss the impact of symmetries of the Hamiltonian, i.e.,
conservation laws. In Sec.~\ref{sec:algorithms} we give concrete
matrix expressions and algorithms for computing the effective
Hamiltonians, and in the canonical case it is, to the author's
knowledge, previously unknown.  In Sec.~\ref{sec:discussion} we sum up
and briefly discuss the results and possible future projects.

\section{Tools and notation}
\label{sec:tools}

\subsection{Linear spaces and operators}
\label{sec:linear}

We shall use the Dirac notation for vectors, inner products and
operators, in order to make a clear, basis-independent formulation.
By $\CalF$, $\CalG$, etc., we denote (finite dimensional) Hilbert
spaces, and vectors are denoted by kets, e.g., $\ket{\psi}$, as usual.
Our underlying Hilbert space is denoted by $\CalH$, with $n =
\dim(\CalH)$. In general, $n$ is infinite. We shall,
however, assume it to be finite. Our results are still valid in the
infinite dimensional case if $H$ is assumed to have a discrete
spectrum and at least $m$ linearly independent eigenvectors.

We are also given a Hamiltonian $H$, a
linear, Hermitian operator (i.e., $H=H^\dag$) on $\CalH$. Its spectral
decomposition is defined to be
\[ H = \sum_{k=1}^n E_k \ket{\psi_k}\bra{\psi_k}. \]
Thus, $E_k$ and $\ket{\psi_k}$ are the (real) eigenvalues and
(orthonormal) eigenvectors, respectively.

We are also given a subspace $\CalP\subset\CalH$, called the model
space, which in principle is arbitrary. Let $\{\ket{e_k}\}_{k=1}^m$ be
an orthonormal basis, for definiteness, viz,
\[ \CalP := \Span \{ \ket{e_k} \::\:k=1,\cdots,m\}. \]
Let $P$ be its orthogonal projector, i.e.,
\[ P : \CalH \rightarrow \CalP, \quad P = \sum_{j=1}^m \ket{e_j}\bra{e_j}, \quad m = \dim(\CalP) \leq n, \]
The basis $\{\ket{e_j}\}_{j=1}^m$ is
commonly taken to be eigenvectors for $H_0$.

The orthogonal complement of the model space, $\CalQ = \CalP^\bot$, has
the orthogonal projector $Q = 1 - P$, and is called
the excluded space.

This division of $\CalH$ into  $\CalP$ and $\CalQ$ transfers to operators in $\CalH$. These are in a natural
way split into four parts, viz, for an arbitrary operator $A$,
\begin{equation} A = (P+Q)A(P+Q) = PAP + PAQ + QAP + QAQ,
  \label{eq:blocks} \end{equation}
where $PAP$ maps the model space into itself, $QAP$ maps $\CalP$ into $\CalQ$, and so forth. It is convenient to picture
this in a block-form of $A$, viz,
\[ A = \begin{bmatrix} PAP & PAQ \\ QAP &  QAQ \end{bmatrix}. \]

\subsection{The singular value decomposition}
\label{sec:svd}

A recurrent tool in this work is the singular value decomposition
(SVD) of an operator $A : \CalX \rightarrow \CalY$. Here, $p =
\dim(\CalX)$ and $q = \dim(\CalY)$ are arbitrary. Then there exists
orthonormal bases $\{ \ket{x_k} \}_{k=1}^{p}$ and $\{ \ket{y_k}
\}_{k=1}^{q}$ of $\CalX$ and $\CalY$, respectively, and
$r=\min(p,q)$ non-negative real numbers $\sigma_k$ with
$\sigma_k\geq \sigma_{k+1}$ for all $k$, such that
\[ A = \sum_{k=1}^{r} \sigma_k \ket{y_k}\bra{x_k}. \]
This is the singular value decomposition (SVD) of $A$, and it always exists.
It may happen that some of the basis vectors do not
participate in the sum; either if $p \neq q$, or if $\sigma_k=0$
for some $k$. 

The vectors $\ket{x_k}$ are called right singular vectors, while
$\ket{y_k}$ are called left singular vectors. The values $\sigma_k$
are called singular values, and $A$ is one-to-one and onto (i.e.,
nonsingular) if and only if $\sigma_k>0$ for all $k$, and $p=q$.
The inverse is then
\[ A^{-1} = \sum_{k=1}^{r}
\frac{1}{\sigma_k}\ket{x_k}\bra{y_k}, \]
as easily verified.

A recursive variational characterization of the singular values and
vectors is the following \cite{Bjorck1973}:
\begin{eqnarray}
  \sigma_k &=&
  \max_{\doubleunder{\ket{u}\in\CalX,\;\braket{u|u}=1}{\braket{u|u_j}=0,\;j<k}} \; \max_{\doubleunder{\ket{v}\in\CalY,\;\braket{v|v}=1}{\braket{v|v_j}=0,\;j<k}}
  \Real \braket{v|A|u} \notag \\ &=:& 
  \braket{v_k|A|u_k}. \label{eq:svd-minimax} 
\end{eqnarray}
The latter equality implicitly states that the maximum is actually
real.  The SVD is very powerful, as it gives an interpretation and
representation of \emph{any} linear operator $A$ as a simple scaling
with respect to one orthonormal basis, and then transformation to
another. The singular vectors are not unique, but the singular values
are.

\subsection{Principal angles and vectors}
\label{sec:pca}

Important tools for comparing linear subspaces $\CalF$ and $\CalG$ of
$\CalH$ are the principal angles and principal
vectors \cite{Knyazev2002,Golub1989}.  
The principal angles generalize the notion
of angles between vectors to subspaces in a natural way. They are also called
canonical angles. Assume that
\[ p = \dim(\CalF) \geq q =\dim(\CalG) \geq 1. \] 
(If $p<q$, we simply exchange $\CalF$ and $\CalG$.) Then, $q$
principal angles $\theta_k\in[0,\pi/2]$, with $\theta_k\leq
\theta_{k+1}$ for all $k$, and the left and right principal vectors
$\ket{\xi_k}\in\CalF$ and $\ket{\eta_k}\in\CalG$
are defined recursively through
\begin{eqnarray}
  \cos\theta_k &=& \max_{\doubleunder{\ket{\xi}\in\CalF,\;\braket{\xi|\xi}=1}{\braket{\xi|\xi_j}=0,\;j<k}} \; \max_{\doubleunder{\ket{\eta}\in\CalG,\;\braket{\eta|\eta}=1}{\braket{\eta|\eta_j}=0,\;j<k}}
  \Real \braket{\xi|\eta} \notag \\ &=:& 
  \braket{\xi_k|\eta_k}. \label{eq:pca-minimax} 
\end{eqnarray}
Again, the last equality implicitly states that the maximum actually
is real. One sees that $\theta_k$ is the
angle between $\ket{\xi_k}\in\CalF$ and $\ket{\eta_k}\in\CalG$. 

It is evident from Eqns.~(\ref{eq:svd-minimax}) and
(\ref{eq:pca-minimax}) that the principal angles and vectors are closely
related to the SVD. Indeed, if we consider the product of the
orthogonal projectors $P_\CalF$ and $P_\CalG$ and compute the SVD, we obtain
\[ P_\CalF P_\CalG = \sum_{k=1}^p \ket{\xi_k}\bra{\xi_k} \sum_{j=1}^q
\ket{\eta_k}\bra{\eta_k} = \sum_{k=1}^q \cos \theta_k
\ket{\xi_k}\bra{\eta_k}, \] where we extended the orthonormal vectors
$\{\ket{\xi_k}\}_{k=1}^q$ with $p-q$ vectors into a basis for $\CalF$,
which is always possible. This equation in particular implies the
additional orthogonality relation $\braket{\xi_j|\eta_k} =
\delta_{j,k}\cos\theta_k$ on the principal vectors.

The principal vectors constitute orthonormal bases that should be
rotated into each other if the spaces were to be aligned. Moreover,
the rotations are by the smallest angles possible.

\section{Effective Hamiltonians}
\label{sec:effham}

\subsection{Similarity transforms}
\label{sec:sim}

The goal of the effective Hamiltonian is to reproduce exactly $m$ of
the eigenvalues, and (necessarily) approximately $m$ of the
eigenvectors. We shall assume that the first $m$ eigenpairs
$(E_k,\ket{\psi_k})$, $k=1,\ldots,m$, defines these. We define the
space $\CalE$ as 
\[ \CalE := \Span \{ \ket{\psi_k} \;:\; k=1,\cdots,m \}. \]
The orthogonal projector $P'$ onto $\CalE$ is
\begin{equation} 
  P' = \sum_{k=1}^m \ket{\psi_k}\bra{\psi_k}.
  \label{eq:eigenspace} \end{equation}

We denote by $(E_k, \ket{\psi_k^\text{eff}})$, $k=1,\ldots,m$, the
effective Hamiltonian eigenvalues and eigenvectors. Of course, the
$\ket{\psi_k^\text{eff}}\in\CalP$ must constitute a basis for $\CalP$,
but not necessary an orthonormal basis. Geometrically, we want
$\ket{\psi^\eff_k}$ to be as close as possible to $\ket{\psi_k}$,
i.e., we want $\CalE$ to be as close to $\CalP$ as possible.

Let $\ket{\widetilde{\psi_k^\text{eff}}}$ be the bi-orthogonal basis,
i.e., $\braket{\widetilde{\psi_j^\eff}|\psi_k^\text{eff}} = \delta_{j,k}$,
so that
\[ P = \sum_{k=1}^m \ket{e_k}\bra{e_k} = \sum_{k=1}^m
\ket{\psi_k^\text{eff}}\bra{\widetilde{\psi_k^\text{eff}}}. \]
The spectral decomposition of $\Heff$ becomes
\[ \Heff = \sum_{k=1}^m E_k
\ket{\psi_k^\text{eff}}\bra{\widetilde{\psi_k^\text{eff}}}. \]
Since $\Heff$ is to have eigenvalues identical to $m$ of those of $H$,
and since $\Heff$ operates only in $\CalP$, we may relate $\Heff$ to
$H$ through a similarity transform, viz,
\begin{equation} \Heff = P\tilde{H}P = P(e^{-S}He^{S})P, \label{eq:1}
\end{equation} where $\exp(S)\exp(-S)=I$. Any
invertible operator has a logarithm, so Eqn.~(\ref{eq:1}) is
completely general.

Now, $\Heff = P\tilde{H}P$ is
an effective Hamiltonian only if the Bloch equation 
\begin{equation} Q\tilde{H}P = Qe^{-S}He^SP = 0 \label{eq:2}
\end{equation} is satisfied \cite{Lindgren1974}, since $\CalP$ is then
invariant under the action of $\tilde{H}$.  The eigenvectors of
$\Heff$ are now given by
\begin{equation} \ket{\psi_k^\text{eff}} = e^{-S} \ket{\psi_k} \in
  \CalP,\quad k=1,\cdots,m. \label{eq:psi-eff-general} \end{equation} Thus, an effective
Hamiltonian can now be defined for every $S$ such that
Eqn.~\eqref{eq:2} holds. It is readily seen that
$\tilde{H}=\tilde{H}^\dag$ if and only if $S$ is skew-Hermitian, i.e.,
that $S^\dag = -S$.  There is still much freedom in the choice of
exponent $S$. Indeed, given any invertible operator $A$ in $\CalP$,
$A^{-1}\Heff A$ is a new effective Hamiltonian with the same effective
eigenvalues as $\Heff$, and $\ket{\psi^\text{eff}} =
A^{-1}\exp(-S)\ket{\psi}$.

\subsection{Geometry of the model space}
\label{sec:geometry}

We will benefit from a detailed discussion of the spaces $\CalE$ and
$\CalP$ before we discuss the Bloch-Brandow and the canonical effective
Hamiltonians in detail.

Since $\dim(\CalP)=\dim(\CalE)=m$, the closeness of the effective and
exact eigenvectors can be characterized and measured by the
orientation of $\CalE$ relative to $\CalP$ in $\CalH$, using $m$
canonical angles $\theta_k$ and principal vectors
$\ket{\eta_k}\in\CalE$ and $\ket{\xi_k}\in\CalP$.
Recall, that $\cos\theta_k = \braket{\xi_k|\eta_k}$ and that the angles
$\theta_k\in[0,\pi/2]$ were the smallest possible such that the
principal vectors are the orthonormal bases of $\CalP$ and $\CalE$
that are closest to each other.

\begin{figure}
\begin{center}
\includegraphics{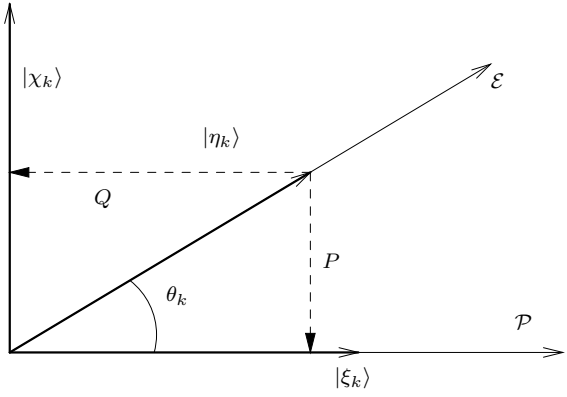}
\caption{Plane spanned by $\ket{\xi_k}$ and $\ket{\eta_k}$. Action of
  projectors $P$ and $Q$ on $\ket{\eta_k}$ indicated\label{fig:rotation}}
\end{center}
\end{figure}

We now define the unitary operator $Z=\exp(G)$ that rotates $\CalP$ into $\CalE$
according to this description, i.e., we should have $Z\ket{\xi_k} =
\ket{\eta_k}$. 
In Fig.~\ref{fig:rotation} the plane spanned by $\ket{\eta_k}$ and
$\ket{\xi_k}$ if $\theta_k>0$ is depicted. 
Recall, that $\braket{\xi_j|\eta_k} = \cos\theta_j\delta_{j,k}$.
Note that $\ket{\xi_k}=\ket{\eta_k}$ if and only if $\theta_k=0$, and
  the plane degenerates into a line. If
  $\theta_k>0$, the vector $\ket{\chi_k}$ is defined so that it together with
$\ket{\xi_k}$ is an orthonormal basis for the plane, viz,
\begin{equation} \ket{\eta_k} = P\ket{\eta_k} + Q\ket{\eta_k} =
\cos(\theta_k)\ket{\xi_k} + \sin(\theta_k)\ket{\chi_k},
\label{eq:chi2} \end{equation}
where 
\begin{equation} \ket{\chi_k} =
  \frac{Q\ket{\eta_k}}{\braket{\eta_k|Q|\eta_k}^{1/2}}. \label{eq:chi}
\end{equation}
Thus, $\{\ket{\chi_k}\}\cup\{\ket{\xi_k}\}$ is an orthonormal basis
for $\CalP\oplus\CalE$, whose dimension is $2m - n_z$, where $n_z$ is
the number of $\theta_k=0$. The set
$\{\ket{\chi_k}:\theta_k>0,\:k=1,\cdots,m\}$ is an orthonormal basis for
$Q\CalE$ which contains $Q\ket{\psi_k}$ for all $k=1,\cdots,m$.

The operator $Z$ is now defined as a rotation in $\CalP\oplus\CalE$,
i.e., by elementary trigonometry,
\begin{eqnarray}
  Z\ket{\xi_k} &:=& \ket{\eta_k} \notag \\
  Z\ket{\eta_k} &:=& 2\cos\theta_k\ket{\eta_k} - \ket{\xi_k} \label{eq:direct-rot-def1}
\end{eqnarray}
In terms of the orthonormal basis, we obtain a manifest planar
rotation for each $k$, i.e.,
\begin{eqnarray}
  Z\ket{\chi_k} &=&  \cos\theta_k\ket{\chi_k}
  -\sin\theta_k\ket{\xi_k} \notag \\
  Z\ket{\xi_k} &=& \sin\theta_k\ket{\chi_k} +\cos\theta_k\ket{\xi_k}. \label{eq:z-ortho2}
\end{eqnarray}
On the rest of the Hilbert space,
$\CalH\ominus(\CalP\oplus\CalE)$, $Z$ is the identity.
The operator $Z$ implements the so-called direct rotation \cite{Davis1970} of $\CalP$ into $\CalE$. 
From Eqn.~(\ref{eq:z-ortho2}) we obtain
\begin{eqnarray}
Z &:=& I + \sum_{k=1}^m [\cos(\theta_k)-1] ( \ket{\chi_k}\bra{\chi_k} +
\ket{\xi_k}\bra{\xi_k} ) \notag \\ & & + \sum_{k=1}^m \sin(\theta_k) ( \ket{\chi_k}\bra{\xi_k} -
\ket{\xi_k}\bra{\chi_k} ),
\label{eq:direct-rot-def}
\end{eqnarray}

It is instructive to exhibit the Lie algebra element $G\in su(n)$ such
that $Z = \exp(G)\in SU(n)$. Since we have Eqn.~\eqref{eq:z-ortho2},
is is easy to do this. Indeed, taking the exponential of
\begin{equation} 
  G = \sum_{k=1}^m \theta_k (
  \ket{\chi_k}\bra{\xi_k} - \ket{\xi_k}\bra{\chi_k}), \label{eq:G} 
\end{equation}
by summing the series for $\sin(\theta)$ and $\cos(\theta)$, we
readily obtain $Z = \exp(G)$, the desired
result. Moreover, observe that the $k$'th term in Eqn.~(\ref{eq:G}) commutes with the
$j$'th term, so, $\exp(G)$ is exhibited as a sequence of commuting rotations using the
canonical angles $\theta_k$.

\subsection{The Bloch-Brandow effective Hamiltonian and the decoupling
operator}

For reference, we review some properties of the Bloch-Brandow
effective Hamiltonian, which we denote by $\Heff^\text{BB}$
\cite{Bloch1958,Brandow1967,Ellis1977,Dean2004}. 
The effective eigenvectors $\ket{\psi_k^\text{eff}}$ are defined by
\begin{equation} \ket{\psi_k^\text{eff}} := P\ket{\psi_k} :=
  \ket{P\psi_k}. \label{eq:bb-effective-eigen} \end{equation} Since
$\ket{P\psi_k}$ are the orthogonal projections of $\ket{\psi_k}$ onto
$\CalP$, we deduce that the Bloch-Brandow effective eigenvectors are
\emph{the closest possible to the exact model space eigenvectors}. In
this sense, the Bloch-Brandow effective Hamiltonian is the
optimal choice.

It is obvious that $\Heff^\text{BB}$ is
non-Hermitian, as rejecting the excluded space eigenvector components
renders the effective eigenvectors non-orthonormal, i.e.,
\[ \braket{\psi^\text{eff}_j|\psi^\text{eff}_k} = \delta_{j,k} -
\braket{\psi_j|Q|\psi_k} \neq \delta_{j,k}. \]
In terms of similarity transforms, we obtain $\Heff^\text{BB}$ by
setting $S = \omega$, the so-called de-coupling operator or
correlation operator \cite{Suzuki1982,Dean2004}. It is
defined by $\omega = Q\omega P$ and the equation
\begin{equation} 
  \omega P\ket{\psi_k} := Q\ket{\psi_k}. \label{eq:omega} 
\end{equation} 
Again, for this to be a meaningful definition,
$\{\ket{P\psi_k}\}_{k=1}^m$ must be a basis for $\CalP$.  

Since $\omega^2 = 0$, $\exp(\pm\omega) = 1 \pm \omega$, and
Eqn.~(\ref{eq:psi-eff-general}) becomes
\[ e^{-\omega}\ket{\psi_k} = (1 - \omega)\ket{\psi_k} =
(1-Q)\ket{\psi_k} = \ket{P\psi_k}. \]
For $\Heff^\text{BB}$ we thus obtain
\begin{equation}
  \Heff^\text{BB} = Pe^{-\omega}He^{\omega} P = PH(P+\omega). \label{eq:heff-bb}
\end{equation}

After this initial review, we now relate $\omega$ to the geometry of
$\CalE$ and $\CalP$. The SVD of $\omega$ is readily obtainable by
expanding the principal vectors $\{\ket{\eta_k}\}_{j=1}^m$ in the $m$
eigenvectors $\{\ket{\psi_k}\}_{k=1}^m$, sets which both constitute a
basis for $\CalE$, and inserting in Eqn.~(\ref{eq:omega}). We have
\begin{eqnarray*} 
  Q \ket{\eta_k} &=& \sum_{j=1}^m Q\ket{\psi_j}\braket{\psi_j|\eta_k}  \\
    &=& \sum_{j=1}^m \omega P \ket{\psi_j}\braket{\psi_j|\eta_k} = \omega P
      \ket{\eta_k},
   \end{eqnarray*}
that is,
\[ \omega \left( \cos\theta_k \ket{\xi_k} \right) = \sin\theta_k
\ket{\chi_j}. \]
The result is
\begin{equation} 
  \omega = \sum_{k=1}^m \tan\theta_k
  \ket{\chi_k}\bra{\xi_k}, \label{eq:omega-svd} 
\end{equation}
which is the SVD of $\omega$. The operator $\omega$ is thus exhibited
as an operator intimately related to the principal angles and vectors
of $\CalP$ and $\CalE$: It transforms the principal vectors of $\CalP$
into an orthonormal basis for $Q\CalE$, with coefficients determined by the
canonical angles $\theta_k$. Using Eqn.~(\ref{eq:chi2}) we obtain an
alternative expression, viz,
\begin{equation} 
  \omega + P = \sum_{k=1}^m \frac{1}{\cos\theta_k}
  \ket{\eta_k}\bra{\xi_k}. \label{eq:P+omega-svd} 
\end{equation}

\subsection{The canonical effective Hamiltonian}

Hermitian effective Hamiltonians have independently been introduced by
various authors since 1929, when Van Vleck
\cite{VanVleck1929,Jordahl1934,Kemble1937} introduced a unitary
transformation $\tilde{H}=\exp(-S)H\exp(S)$ to decouple the model
space to second order in the interaction. In 1963, Primas
\cite{Primas1963} considered an order by order expansion of this
$\tilde{H}$ using the Baker-Campbell-Hausdorff formula and commutator
functions to determine $S$, a technique also used in many other settings in which a
transformation is in a Lie group, see, e.g., Ref.~\cite{Blanes1998}
and references therein. This approach was elaborated by Shavitt and
Redmon \cite{Shavitt1980}, who were the first to mathematically
connect this Hermitian effective Hamiltonian to $\Heff^\text{BB}$, as
in Eqn.~\eqref{eq:G-omega} below. In the nuclear physics community,
Suzuki \cite{Suzuki1982a} has been a strong advocate of Hermitian
effective interactions and the $a$-body sub-cluster approximation to
the $A$-body effective interaction
\cite{Suzuki1982a,Suzuki1982,Navratil2000}.  Hermiticity in this case
is essential.

Even though a Hermitian effective Hamiltonian is not unique due to the
non-uniqueness of $S=-S^\dag$, the various Hermitian effective
Hamiltonians put forward in the literature  all turn
out to be equivalent \cite{Klein1974}. In the spirit of
Klein and Shavitt \cite{Klein1974,Shavitt1980} we employ the term
``canonical effective Hamiltonian'' since this emphasizes the
``natural'' and geometric nature of the Hermitian effective Hamiltonian, which we
denote by $\Heff^\text{c}$.

Recall the spectral decomposition 
\[ \Heff^\text{c} = \sum_{k=1}^m E_k \ket{\psi_k^\text{eff}}
\bra{\psi_k^\text{eff}}, \] where the (orthonormal) effective
eigenvectors are now defined by the following optimization
property: \emph{The effective eigenvectors $\ket{\psi_k^\eff}$ are the
  closest possible to the exact eigenvectors $\ket{\psi_k}$
  while still being orthonormal.} Thus, where the Bloch-Brandow
approach \emph{globally} minimizes the distance between the
eigenvectors, at the cost of non-orthonormality, the canonical approach
has the unitarity constraint on the similarity transformation,
rendering $\Heff^\text{c}$ Hermitian. 

Given a collection $\Phi = \{ \ket{\phi_1}, \ldots,
\ket{\phi_m}\}\subset\CalP$ of $m$ vectors, which are candidates for
effective eigenvectors, define the functional $S[\Phi]$ by
\begin{eqnarray} 
  S[\Phi] &:=& \sum_{k=1}^m \| \ket{\phi_k} -
  \ket{\psi_k} \|^2 \notag \\
  &=& K + \sum_{k=1}^m \|
  P\ket{\psi_k} - \ket{\phi_k} \|^2 \notag \\
  &=& K + \sum_{k=1}^m
  \|\ket{\phi_k}\|^2 - 2\Real \sum_{k=1}^m \braket{\psi_k|P|\phi_k}, 
  \label{eq:functional}
\end{eqnarray} 
where $K=\sum_{k=1}^m \|Q\ket{\psi_k}\|^2$ is a constant in this
context. The effective eigenvectors are now minimizers of $S[\Phi]$.

The global minimum, when $\Phi\subset\CalP$ is allowed to vary freely,
is attained for $\ket{\phi_k} = \ket{P\psi_k}$, the
Bloch-Brandow effective eigenvectors. However, the canonical effective
eigenvectors are determined by minimizing $S[\Phi]$ over all orthonormal
sets $\Phi$, which then becomes equivalent to maximizing the last term
in Eqn.~(\ref{eq:functional}), i.e., the overlaps
$\sum_k \Real \braket{\psi_k|P|\phi_k}$ under the orthonormality constraint.

We will now prove the striking fact that the solution is given by
\begin{equation} \ket{\psi_k^\eff} = \ket{\phi_k} = e^{-G}
  \ket{\psi_k}, \label{eq:psi-eff-canonical} \end{equation}
where the unitary operator $Z := \exp(G) \in SU(n)$ is the
rotation~(\ref{eq:direct-rot-def}). Equation
(\ref{eq:psi-eff-canonical}) should be compared with
Eqn.~(\ref{eq:psi-eff-general}). Thus,
the exact eigenvectors are simply the \emph{direct rotations of
  the effective eigenvectors from the model space into $\CalE$}.

Let us expand $\ket{\psi_k}\in\CalE$ and $\ket{\phi_k}\in \CalP$ in the
principal vector bases, viz,
\begin{eqnarray*}
  \ket{\psi_k} &=& \sum_{j=1}^m
  \ket{\eta_j}\braket{\eta_j|\psi_k}, \\
  \ket{\phi_k} &=& \sum_{j=1}^m
  \ket{\xi_j}\braket{\xi_j|\phi_k} .
\end{eqnarray*}
Using $\braket{\eta_j|\xi_k} = \delta_{j,k}\cos\theta_j$,
we compute the sum $A := \sum_{k=1}^m \braket{\psi_k|P|\phi_k}$ as
\begin{eqnarray*}
  A &=& \sum_{k,j,\ell=1}^m
  \braket{\psi_k|\eta_j}\braket{\eta_j|\xi_\ell}\braket{\xi_\ell|\phi_k}
  \\
  &=& 
  \sum_{j,k=1}^m\cos\theta_j \braket{\xi_j|\phi_k}\braket{\psi_k|\eta_j}
\end{eqnarray*}
Now, 
\[ A = \sum_{j=1}^m \cos\theta_j u_{j,j}, \]
where $u_{j,k}$ is a unitary matrix, which
implies $|u_{j,j}|\leq 1$. Moreover, $u_{j,j}=1$
for all $j$
if and only if $u_{j,k}=\delta_{j,k}$, which then maximizes $A$, and also $\Real A$. Thus,
\[ \sum_{k=1}^m \braket{\xi_j|\phi_k}\braket{\psi_k|\eta_\ell} =
\delta_{j,\ell}, \]
i.e.,
\[ \braket{\xi_j|\phi_k} =
\braket{\eta_j|\psi_k} = \braket{\xi_j|Z^\dag|\psi_k}, \]
from which Eqn.~(\ref{eq:psi-eff-canonical}) follows since
$\{\ket{\xi_k}\}_{k=1}^m$ is a basis for $\CalP$,
and the proof is complete. 

The similarity transform in Eqn.~(\ref{eq:1}) is thus manifest, with $S=G$, viz,
\begin{equation} \Heff^\text{c} = P Z^\dag H Z P = P e^{-G} H e^{G}
  P. \label{eq:effham-sim} \end{equation}
Moreover, $Q\tilde{H}P=P\tilde{H}Q=0$, verifying that the direct rotation in fact block
diagonalizes $H$.

\subsection{Computing $\ket{\psi_k^\eff}$}

Assume that $\ket{P\psi_k}:=P\ket{\psi_k}$, $k=1,\cdots,m$
are available. The effective eigenvectors $\ket{\psi_k^\eff}$ are then given
by a basis change $F$, i.e., the operator
$F:\CalP\rightarrow\CalP$ defined by
\[ F\ket{P\psi_k} := \ket{\psi_k^\text{eff}}. \]
Using the principal vector basis we obtain
\begin{eqnarray*}
  F\ket{P\psi_k} &=& F P\sum_{j=1}^m \ket{\eta_j}\braket{\eta_j|\psi_k} \\
  &=& F\sum_{j=1}^m \cos\theta_j
  \ket{\xi_j}\braket{\eta_j|\psi_k} \\
  &:=& \sum_{j=1}^m \ket{\xi_j}\braket{\xi_j|\psi_k^\text{eff}} \\ 
  &=& \sum_{j=1}^m \ket{\xi_j}\braket{\eta_j|\psi_k} 
\end{eqnarray*}
from which we get the SVD
\begin{eqnarray}
  F &:=& \sum_{k=1}^m \frac{1}{\cos\theta_k}\ket{\xi_k}\bra{\xi_k}
  \notag \\
  &=& (\omega^\dag\omega + P)^{1/2}, 
  \label{eq:F}
\end{eqnarray}
where we have used Eqn.~(\ref{eq:P+omega-svd}).
From Eqn.~(\ref{eq:F}) we see that $F$ is symmetric and positive
definite. Moreover, smaller angles $\theta_k$ means $F$ is closer to
the identity, consistent with $\CalE$ is closer to $\CalP$.

Let $\ket{P\psi_k}$ now be given in the orthonormal ``zero order'' basis
$\{\ket{e_k}\}_{k=1}^m$ for $\CalP$, i.e., we have the basis change
operator $\tilde{U}$ given by
\begin{equation} \tilde{U} := \sum_{k=1}^m \ket{P\psi_k}\bra{e_k},
  \label{eq:U} \end{equation}
which transforms from the given basis to the Bloch-Brandow effective
eigenvectors. In terms of the principal vector basis,
\begin{eqnarray}
  \tilde{U} &=& \sum_{j=1}^m \cos\theta_j \ket{\xi_j}\bra{\eta_j} \sum_k
  \ket{\psi_k}\bra{e_k} \notag \\
  &=:& \sum_{j=1}^m \cos\theta_j \ket{\xi_j}\bra{y_j}, \label{eq:U-svd}
\end{eqnarray} 
which is, in fact, the SVD since the last sum
over $k$ is a unitary map from $\CalP$ to $\CalE$. In the
operator $\tilde{U}\tilde{U}^\dag$ this basis-dependent factor cancels, viz,
\begin{eqnarray*}
  \tilde{U}\tilde{U}^\dag &=& \sum_{k=1}^m \ket{P\psi_k}\bra{P\psi_k} \\ 
  &=& \sum_{k=1}^m \cos^2\theta_k \ket{\xi_k}\bra{\xi_k},
\end{eqnarray*}
that is,
\[ F = (\tilde{U}\tilde{U}^\dag)^{-1/2}. \]
If we seek $\ket{\psi_k^\eff}$ in the basis $\{\ket{e_k}\}_{k=1}^m$ as
well, we let $\tilde{V}$ be the corresponding basis change
operator, i.e.,
\begin{equation} \tilde{V} := F\tilde{U} =
  (\tilde{U}\tilde{U}^\dag)^{-1/2} \tilde{U}. \label{eq:V} \end{equation}
Equation (\ref{eq:V}) shows that $\ket{\psi_k^\text{eff}}$ is obtained
by ``straightening out'' $\ket{P\psi_k}$, and that this depends
\emph{only} on the latter vectors. 
This is, in fact, an alternative to
the common Gram-Schmidt orthogonalization used in mathematical
constructions and proofs. It was first introduced by L\"owdin
\cite{Lowdin1950} under the name ``symmetric orthogonalization'', and
so-called ``L\"owdin bases'' are widely-used in quantum chemistry,
where non-orthogonal basis functions are orthogonalized according to
Eqn.~(\ref{eq:V}). It seemingly requires both inversion and matrix square
root, but is easily computed using the SVD. Combining
Eqns.~(\ref{eq:F}) and (\ref{eq:U-svd}) gives
\begin{equation} \tilde{V} = \sum_{k=1}^m
\ket{\xi_k}\bra{y_k}, \label{eq:V-svd}
\end{equation}
so that if the SVD (\ref{eq:U-svd}) is available, $\tilde{V}$ is
readily computed. Eqn.~(\ref{eq:V-svd}) is easily expressed in
terms of matrices, but we defer the discussion to
Sec.~\ref{sec:algorithms}.

\subsection{Shavitt's expression for $\exp(G)$}

Shavitt and Redmon \cite{Shavitt1980} proved that 
\begin{equation} G = \tanh^{-1} (\omega - \omega^\dag)
  \label{eq:G-omega} \end{equation}
gives the Lie algebra element for the unitary operator
$Z=\exp(G)$. The quite complicated proof was done using an
expansion of the similarity transform using the
Baker-Campbell-Hausdorff formula.

It may be clear now, that in the present context we obtain the result
simply as a by-product of the treatment in Section \ref{sec:geometry}
and the SVD (\ref{eq:omega-svd}) of $\omega$, given in terms of the
principal vectors and angles. We prove this here.

The function $\tanh^{-1}(z)$ is defined by its (complex) Taylor
expansion about the origin, i.e.,
\begin{equation} \tanh^{-1} (z) = \sum_{n=0}^\infty \frac{z^{2n+1}}{2n+1}.\label{eq:artanh-series} \end{equation}
The series converges for $|z|<1$.
Moreover,
\begin{equation} \tanh^{-1} (z) = \frac{1}{2}\ln
  \left(\frac{1+z}{1-z}\right), \label{eq:artanh} \end{equation}
also valid for $|z|<1$. For $z
:= \omega-\omega^\dag$ we compute
\[ z = \sum_{k=1}^m \mu_k (\ket{\chi_k}\bra{\xi_k} -
\ket{\xi_k}\bra{\chi_k}), \quad \mu_k := \tan(\theta_k). \]
Using orthogonality relations between $\ket{\xi_k}$ and $\ket{\chi_k}$
we obtain
\begin{eqnarray*}
 z^{2n+1} &=& (-1)^n \sum_{k=1}^m \mu_k^{2n+1} (\ket{\chi_k}\bra{\xi_k} -
\ket{\xi_k}\bra{\chi_k}) \\
&=& i \sum_{k=1}^m (-i\mu_k)^{2n+1} (\ket{\chi_k}\bra{\xi_k} -
\ket{\xi_k}\bra{\chi_k}).
\end{eqnarray*}
Using $i\tanh^{-1}(-iz) = \tan^{-1}(z)$, we sum the series
(\ref{eq:artanh-series}) to
\[ G = \tanh^{-1} (z) = \sum_{k=1}^m \theta_k (
\ket{\chi_k}\bra{\xi_k} - \ket{\xi_k}\bra{\chi_k} ), \] which is
identical to Eqn.~(\ref{eq:G}). The series does not converge for
$\theta_k\geq \pi/4$, but the result is trivially analytically
continued to arbitrary $0\leq\theta_k\leq\pi/2$.

We now turn to the effective Hamiltonian. It is common
\cite{Suzuki1982,Navratil2000,Caurier2005} to compute $\Heff^\text{c}$ in
terms of $\omega$ directly, using the definition \eqref{eq:artanh} of
$\tanh^{-1}(z)$, which implies
\[ e^{\pm\tanh^{-1}(z)} = \sqrt{\frac{1\pm z}{1\mp z}} = \frac{1\pm
  z}{\sqrt{1-z^2}}, \]
Upon insertion into $\tilde{H} = \exp(-G)H\exp(G)$, we obtain
\[ \tilde{H} = \frac{1 - \omega + \omega^\dag}{\sqrt{1 +
    \omega^\dag\omega + \omega\omega^\dag}} H \frac{1 + \omega - \omega^\dag}{\sqrt{1 +
    \omega^\dag\omega + \omega\omega^\dag}}.  \]
Projecting onto $\CalP$, the effective Hamiltonian becomes
\begin{equation} \Heff = (P+ \omega^\dag \omega)^{-1/2}(P+\omega^\dag)
H (P + \omega )(P + \omega^\dag \omega )^{-1/2}.
\label{eq:heff-old1} \end{equation}
By using the Bloch-equation (\ref{eq:2}) for the Bloch-Brandow
effective Hamiltonian, we
may eliminate $QHQ$ from the above expression for $\Heff$, yielding
\begin{equation} \Heff^\text{c} = (P + \omega^\dag\omega )^{1/2} H ( P
  +\omega ) (P + \omega^\dag\omega )^{-1/2}. \label{eq:effham-old}
\end{equation} 
This expression is commonly implemented in numerical applications \cite{Navratil2000,Hagen2006}. By comparing with
Eqns.~(\ref{eq:F}) and (\ref{eq:heff-bb}) we immediately see that
\begin{equation} \Heff^\text{c} = F\Heff^\text{BB} F^{-1}, \label{eq:effham-old2}
\end{equation} 
which gives $\Heff^\text{c}$ as a similarity transform of
$\Heff^\text{BB}$.  In themselves, Eqns.~(\ref{eq:effham-old}) and
(\ref{eq:effham-old2}) are not manifestly Hermitian, stemming from the
elimination of $QHQ$.  An implementation would require complicated
matrix manipulations, including a matrix square root. It is therefore
better to compute $\Heff^\text{c}$ using
\[ \Heff^\text{c} = \sum_k E_k
\ket{\psi^\text{eff}_k}\bra{\psi^\text{eff}_k} \] together with
Eqn.~(\ref{eq:V-svd}), where the most complicated operation is the SVD
of the operator $\tilde{U}$ given by Eqn.~(\ref{eq:U}). In
Sec.~\ref{sec:algorithms} we give a concrete matrix expression for
$\Heff^\text{c}$.

\subsection{Commuting observables}
\label{sec:commuting-observables}

Great simplifications arise in the general quantum problem if
continuous symmetries of the Hamiltonian can be identified, i.e., if one can find
one or more observables $S$ such that $[H,S]=0$. Here, we discuss the
impact of such symmetries of $H$ on the effective Hamiltonian $\Heff$;
both in the Bloch-Brandow and the canonical case. We point out the
importance of choosing a model space that is an invariant of $S$ as
well, i.e., $[S,P]=0$. In fact, we prove that this is the case if and
only if $[\Heff,S]=0$, i.e., $\Heff$ has the same continuous symmetry.

Let $S=S^\dag$ be an observable such that $[H,S]=0$, i.e., $H$ and $S$
have a common basis of eigenvectors. We shall assume that
$\{\ket{\psi_k}\}_{k=1}^n$ is such a basis, viz,
\begin{eqnarray}
  H\ket{\psi_k} &=& E_k\ket{\psi_k}, \notag \\
  S\ket{\psi_k} &=& s_k\ket{\psi_k}. \label{eq:common-basis}
\end{eqnarray}
In general, there will be degeneracies in both $E_k$ and
$s_k$.

We now make the important assumption that 
\begin{equation} 
  [S,P]=0, 
  \label{eq:s-assum} 
\end{equation}
which is equivalent to
\begin{equation*} S = PSP + QSQ. \end{equation*}
Under the assumption (\ref{eq:s-assum}), we have
\[ S\ket{P\psi_k} = PS\ket{\psi_k} = s_k\ket{P\psi_k}, \]
so that the Bloch-Brandow effective eigenvectors are still
eigenvectors of $S$ with the same eigenvalue $s_k$. Moreover, as we
assume that $\{\ket{P\psi_k}\}_{k=1}^m$ is a (non-orthonormal) basis
for $\CalP$, this not possible if $[S,P]\neq 0$. Thus,
$[\Heff^\text{BB},S]=0$ if and only if $[S,P]=0$ (in addition to the
assumption (\ref{eq:common-basis}).)

The assumption (\ref{eq:s-assum}) also implies that $[S,\omega]=0$,
where $\omega=Q\omega P$ is the de-coupling operator. We prove this by
checking that it holds for \emph{all} $\ket{\psi_k}$. For $k\leq m$,
\begin{equation}
  \omega S \ket{\psi_k} = s_k
Q\ket{\psi_k} = S Q \ket{\psi_k} = S\omega
\ket{\psi_k}, \label{eq:omegaS-commute} \end{equation}
while for $k>m$ we need to expand $P\ket{\psi_k}$ in $\ket{P\psi_j}$,
$j\leq m$, viz,
\[ P\ket{\psi_k} = \sum_{j=1}^m
\ket{P\psi_j}\braket{\widetilde{P\psi_j} | P | \psi_k }, \quad k>m \]
and use Eqn.~(\ref{eq:omegaS-commute}). Furthermore,
$[S,\omega^\dag]^\dag=[\omega^\dag,S] = 0$.
It follows that 
\[ [S,(\omega - \omega^\dag)^n] = 0, \quad n = 0,1,\ldots, \]
and, by Eqn.~(\ref{eq:artanh-series}), that
\[ [S,e^G]  = [S,e^{-G}] = 0. \]
This gives
\[ S\ket{\psi_k^\text{eff}} = S e^{-G}\ket{\psi_k} =
s_k\ket{\psi^\text{eff}_k}. \]
Again, since $\{\ket{\psi^\text{eff}_k}\}_{k=1}^m$ is a basis for
$\CalP$, this holds if and only if $[S,P]=0$.
Accordingly, $[\Heff^\text{c},S]=0$ if and only if $[S,P]=0$ (and the
assumption (\ref{eq:common-basis}).)

The importance of this fact is obvious. If one starts with a
Hamiltonian that conserves, say, angular momentum, and computes the
effective interaction using a model space that is \emph{not} an
invariant for the angular momentum operator, i.e., not rotationally
symmetric, then the final Hamiltonian will not have angular momentum
as a good quantum number.

One possible remedy if $[P,S]\neq 0$ is to define the effective observable
$S_\text{eff} := P\exp(-G)S\exp(G)P$ (which in the commuting case is
equal to $PSP$) which obviously commutes with $\Heff$ and satisfies 
\[ S_\text{eff}\ket{\psi_k^\text{eff}} =
s_k\ket{\psi_k^\text{eff}}. \]
This amounts to modifying the concept of rotational symmetry in the
above example.

The assumptions (\ref{eq:common-basis}) and (\ref{eq:s-assum}) have
consequences also for the
structure of the principal vectors $\ket{\xi_k}\in\CalP$ and
$\ket{\eta_k}\in\CalE$. Indeed, write
\begin{eqnarray*}
  \CalE &=& \bigoplus_s \CalE_s \\
  \CalP &=& \bigoplus_s \CalP_s, 
\end{eqnarray*}
where the sum runs over all distinct eigenvalues $s_k$, $k=1,\cdots,m$ of $S$, and where
$\CalE_s$ ($\CalP_s$) is the corresponding eigenspace, i.e.,
\begin{eqnarray*}
  \CalE_s &:=& \Span \left\{ \ket{\psi_k} \; : \; S\ket{\psi_k} =
  s\ket{\psi_k} \right\} \\
  \CalP_s &:=& \Span \left\{ \ket{\psi^\text{eff}_k} \; : \; S\ket{\psi^\text{eff}_k} =
  s\ket{\psi^\text{eff}_k} \right\}.
\end{eqnarray*}
The eigenspaces are all mutually orthogonal, viz, $\CalE_s \bot
\CalE_{s'}$, $\CalP_s \bot \CalP_{s'}$, and $\CalE_s\bot\CalP_{s'}$, for $s\neq s'$. The
definition (\ref{eq:pca-minimax}) of the principal vectors and angles can then we written
\begin{eqnarray*} \cos(\theta_k) &=& \max_s
  \max_{\doubleunder{\ket{\xi}\in\CalP_s,\;\braket{\xi|\xi}=1}{\braket{\xi_j|\xi}=0,\;j<k}} \;
  \max_{\doubleunder{\ket{\eta}\in\CalE_s,\;\braket{\eta|\eta}=1}{\braket{\eta_j|\eta}=0,\;j<k}} \Real \braket{\xi|\eta} \notag  \\ &=:&
  \braket{\xi_k|\eta_k}, 
\end{eqnarray*}
Thus, for each $k$, there is an eigenvalue $s$ of $S$ such that 
\begin{eqnarray*}
  S\ket{\xi_k} &=& s\ket{\xi_k} \\
  S\ket{\eta_k} &=& s\ket{\eta_k}, 
\end{eqnarray*}
showing that the principal vectors are eigenvectors of $S$ if and only
if $[S,P]=0$, $[S,H]=0$, and the assumption (\ref{eq:common-basis}).

The present symmetry considerations imply that model spaces obeying as
many symmetries as possible should be favored over less symmetric
model spaces, since these other model spaces 
become less ``natural'' or ``less effective'' in the sense that their
geometry is less similar to the original Hilbert space. This is most
easily seen from the fact that principal vectors are eigenvectors for
the conserved observable $S$. This may well have great consequences
for the widely-used sub-cluster approximation to the effective
Hamiltonian in no-core shell model calculations
\cite{Klein1974,DaProvidencia1964,Navratil2000}, where one constructs
the effective Hamiltonian for a system of $a$ particles in order to
obtain an approximation to the $A>a$-body effective Hamiltonian. The
model space in this case is constructed in different ways in different
implementations. Some of these model spaces may therefore be better
than others due to different symmetry properties.

\section{Matrix formulations}
\label{sec:algorithms}

\subsection{Preliminaries}
\label{sec:algorithms-pre}

Since computer calculations are invariably done using matrices for
operators, we here present matrix expressions for $\Heff^\text{c}$ and
compare them to those usually programmed in the literature, as well as
expressions for $\Heff^\text{BB}$ and $\omega$.

Recall the standard basis $\{\ket{e_k}\}_{k=1}^n$ of $\CalH$, where
the $\{\ket{e_k}\}_{k=1}^m$ constitute a basis for $\CalP$. These are usually
eigenvectors of the unperturbed ``zero order'' Hamiltonian $H_0$, but
we will not use this assumption. As previously we also assume without loss that the
eigenpairs we wish to approximate in $\Heff$ are
$\{\ket{\psi_k}\}_{k=1}^m$.

An operator $A : \CalH\rightarrow \CalH$ has a matrix
$\mathsf{A}\in\mathbb{C}^{n\times n}$
associated with it. The matrix elements are given by $\mathsf{A}_{jk}
= \braket{e_j|A|e_k}$ such that
\[ A = \sum_{j,k=1}^n \ket{e_j}\bra{e_j}A\ket{e_k}\bra{e_k} =
\sum_{j,k=1}^n \ket{e_j} \mathsf{A}_{jk} \bra{e_k}. \] 
Similarly, any vector
$\ket{\phi}\in\CalH$ has a column vector $\vec{\phi} \in \mathbb{C}^n$ associated
with it, with $\vec{\phi}_j = \braket{e_j|\phi}$.  We will also view
dual vectors, e.g., $\bra{\psi}$, as
row vectors.

The model space $\CalP$ and the excluded space $\CalQ$ are
conveniently identified with $\mathbb{C}^m$ and $\mathbb{C}^{n-m}$,
respectively. Also note that $PAP$, $PAQ$, etc., are identified
with the upper left $m\times m$, upper right $m\times (n-m)$, etc.,
blocks of $\mathsf{A}$ as in Eqn.~(\ref{eq:blocks}).
We use a notation inspired by \textsc{fortran} and \textsc{matlab} and
write
\[ \mathsf{PAP} = \mathsf{A}(\range{1}{m},\range{1}{m}), \quad \mathsf{PAQ} =
\mathsf{A}(\range{1}{m},\range{m+1}{n}), \]
and so forth.

We introduce the unitary operator $U$ as
\[ U = \sum_{k=1}^n \ket{\psi_k}\bra{e_k}, \]
i.e., a basis change from the chosen standard basis to the eigenvector
basis. The columns of $\mathsf{U}$ are the eigenvectors'
components in the standard basis, i.e.,
\[ \mathsf{U}_{jk} = \vec{\psi}_{k,j} = \braket{e_j|\psi_k}, \]
and are typically the eigenvectors returned from a computer
implementation of the spectral decomposition, viz,
\begin{equation} \mathsf{H} = \mathsf{UEU}^\dag,\quad
  \mathsf{E} = \diag(E_1,\ldots,E_n). \label{eq:diagonalization} \end{equation}
The SVD is similarly transformed to matrix form.  The SVD
defined in Sec.~\ref{sec:svd} is then formulated as follows: For any
matrix $\mathsf{A}\in\mathbb{C}^{q\times r}$ there exist matrices
$\mathsf{X}\in\mathbb{C}^{q\times p}$ ($p=\min(q,r)$) and
$\mathsf{Y}\in\mathbb{C}^{r\times p}$, such that $\mathsf{X}^\dag
\mathsf{X} = \mathsf{Y}^\dag \mathsf{Y} = I_p$ (the identity matrix
$\mathbb{C}^{p\times p}$), and a non-negative diagonal matrix
$\mathsf{\Sigma}\in\mathbb{R}^{p\times p}$ such that
\[ \mathsf{A} = \mathsf{X}\mathsf{\Sigma}\mathsf{Y}^\dag. \]
Here, $\mathsf{\Sigma} = \diag(\sigma_1,\cdots,\sigma_p)$, $\sigma_k$ being
the singular values. 

The columns of $\mathsf{X}$ are the left singular vectors' components, i.e.,
$\mathsf{X}_{j,k} = \braket{e_j|x_k}$, and similarly for $\mathsf{Y}$
and the right singular vectors. The difference between the two SVD
formulations is then purely geometric, as the matrix formulation
favorizes the standard bases in $\CalX$ and $\CalY$.

The present version of the matrix SVD is often referred to as the
``economic'' SVD, since the matrices $\mathsf{X}$ and $\mathsf{Y}$ may be
extended to unitary matrices over $\mathbb{C}^q$ and $\mathbb{C}^r$,
respectively, by adding singular values $\sigma_k=0$, $k>m$. The
matrix $\mathsf{\Sigma}$ is then a $q\times r$ matrix with ``diagonal'' given
by $\sigma_k$. This is the ``full'' SVD, equivalent to our basis-free
definition.

\subsection{Algorithms}
\label{sec:algorithms-expr}

Let the $m$ eigenvectors $\ket{\psi_k}$ be calculated and arranged in
a matrix $\mathsf{U}$, i.e., $\psi_k = \mathsf{U}(\range{1}{n},k)$
(where the subscript does \emph{not} pick a single component). Consider the
operator $\tilde{U}$ defined in Eqn.~(\ref{eq:U}), whose matrix'
columns are the Bloch-Brandow effective eigenvectors $\ket{P\psi_k}$
in the standard basis, viz,
\[ \tilde{\mathsf{U}} = \mathsf{U}(\range{1}{m},\range{1}{m}). \]
The columns of the matrix of
$\tilde{V}=(\tilde{U}\tilde{U}^\dag)^{-1/2}\tilde{U}$ are the
canonical effective eigenvectors $\vec{\psi}^\text{eff}_k$. The SVD
(\ref{eq:U-svd}) can be written
\[ \tilde{\mathsf{U}} = \mathsf{X\Sigma Y}^\dag, \]
which gives
\[ \tilde{\mathsf{U}}\tilde{\mathsf{U}}^\dag =
\mathsf{X\Sigma}^2\mathsf{X}^\dag. \]
Since $\mathsf{\Sigma}_{kk} = \cos\theta_k > 0$, we obtain
\[ (\tilde{\mathsf{U}}\tilde{\mathsf{U}}^\dag)^{-1/2} =
\mathsf{X\Sigma}^{-1}\mathsf{X}^\dag, \]
which gives, when applied to $\tilde{\mathsf{U}}$
\[ \tilde{\mathsf{V}} =
(\tilde{\mathsf{U}}\tilde{\mathsf{U}}^\dag)^{-1/2}\tilde{\mathsf{U}} =
\mathsf{XY}^\dag. \] Thus, we obtain the canonical effective
eigenvectors by taking the matrix SVD of $\tilde{\mathsf{U}} =
\mathsf{U}(\range{1}{m},\range{1}{m})$ and multiplying together the
matrices of singular vectors. As efficient and robust SVD
implementations are almost universally available, e.g., in the
\textsc{lapack} library, this makes the canonical effective
interaction much easier to compute compared to
Eqn.~(\ref{eq:effham-old}), viz,
\[ \mathsf{H}_\text{eff}^\text{c} =
\tilde{\mathsf{V}}\mathsf{E}(\range{1}{m},\range{1}{m})\tilde{\mathsf{V}}^\dag. \]
This version requires one SVD computation and three matrix
multiplications, all with $m\times m$ matrices, one of which is diagonal. Equation
(\ref{eq:effham-old}) requires, on the other hand, several more matrix
multiplications, inversions and the square root computation.
The Bloch-Brandow effective Hamiltonian is simply calculated by
\[ \mathsf{H}_\text{eff}^\text{BB} =
\tilde{\mathsf{U}}\mathsf{E}(\range{1}{m},\range{1}{m})\tilde{\mathsf{U}}^{-1}. \]

For the record, the matrix of $\omega$ is given by
\[ \bm{\omega} = \mathsf{U}(\range{m+1}{n},\range{1}{m}) \tilde{\mathsf{U}}^{-1}, \]
although we have no use for it when using the SVD based algorithm. It
may be useful, though, to be able to compute the principal vectors for
$\CalP$ and $\CalE$. For this, one may compute the SVD of $\omega$ or
of $\mathsf{PP}' =
\tilde{\mathsf{U}}\mathsf{U}(\range{1}{n},\range{1}{m})^\dag$, the latter which
gives $\cos\theta_k$, $\ket{\xi_k}$ and $\ket{\eta_k}$ directly in
the standard basis as singular values and vectors, respectively.

\section{Discussion and and outlook}
\label{sec:discussion}

We have characterized the effective Hamiltonians commonly used in
nuclear shell-model calculations in terms of geometric properties of
the spaces $\CalP$ and $\CalE$. The SVD and the principal angles and
vectors were central in the investigation. While the Bloch-Brandow
effective Hamiltonian is obtained by orthogonally projecting $\CalE$
onto $\CalP$, thereby \emph{globally} minimizing the norm-error of the
effective eigenvectors, the canonical effective Hamiltonian is
obtained by rotating $\CalE$ into $\CalP$ using $\exp(-G)$,
which minimizes the norm-error while retaining orthonormality of the
effective eigenvectors. Moreover, we obtained a complete description
of the de-coupling operator $\omega$ in terms of the principal angles
and vectors defining $\exp(G)$.

An important question is whether the present treatment generalizes to
infinite dimensional Hilbert spaces. Our analysis fits
into the general assumptions in the literature, being that
$n=\dim(\CalH)$ is large but finite, or at least that the spectrum of
$H$ purely discrete. A minimal requirement is that $H$ has $m$
eigenvalues, so that $\CalE$ can be constructed. In particular, the
SVD generalizes to finite rank operators in the infinite dimensional
case, and are thus valid for all the operators considered here even
when $n=\infty$.

Unfortunately, $H$ has almost never a purely discrete spectrum. It is
well-known that the spectrum in general has continuous parts and
resonances embedded in these, and a proper theory should treat these
cases as well as the discrete part. In fact, the treatments of $\Heff$
in the literature invariably glosses over this. It is an interesting
future project to develop a geometric theory for the effective
Hamiltonians which incorporates resonances and continuous spectra.

The geometrical view simplified and unified the available
treatments in the literature somewhat, and offered further insights into
the effective Hamiltonians. Moreover, the the symmetry
considerations in Sec.~\ref{sec:commuting-observables} may have
significant bearing on the analysis of perturbation expansions and
the properties of sub-cluster approximations to $\Heff^\text{c}$.

Indeed, it is easy to see, that if we have a \emph{complete} set of
commuting observables (CSCO) \cite{Messiah1999} for $H_0$, and the
same set of observables form a CSCO for $H_1$, all eigenvalues and
eigenfunctions of $H(z) = H_0 + zH_1$ are analytic in
$z\in\mathbb{C}$, implying that the Rayleigh-Schroedinger perturbation
series for $H = H_0 + H_1$ converges (i.e., at $z=1$)
\cite{Schucan1973}.  Intuitively, the fewer commuting observables we
are able to identify, the more likely it is that there are
singularities in $\Heff(z)$, so called intruder states. The
Rayleigh-Schroedinger series diverges outside the singularity closest
to $z=0$ \cite{Schucan1973}, and in nuclear systems this singularity
is indeed likely to be close to $z=0$. On the other hand, resummation
of the series can be convergent and yield an analytic continuation of
$\Heff$ outside the region of convergence \cite{Schaefer1974}. To the
author's knowledge, there is no systematic treatment of this
phenomenon in the literature. On the contrary, to be able to do such a
resummation is sort of a ``holy grail'' of many-body perturbation
theory. A geometric study of the present kind to many-body
perturbation theory and diagram expansions may yield a step closer to
this goal, as we have clearly identified the impact of commuting
observables on the principal vectors of $\CalE$ and $\CalP$.

We have also discussed a compact algorithm in terms of matrices to
compute $\Heff^\text{c}$, relying on the SVD. To the author's
knowledge, this algorithm is previously unpublished. Since robust and
fast SVD implementations are readily available, e.g., in the
\textsc{lapack} library, and since few other matrix manipulations are
needed, it should be preferred in computer implementations.

As stressed in the Introduction, the algorithms presented are really
only useful if we compute the \emph{exact} effective
Hamiltonian, as opposed to a many-body perturbation theoretical
calculation, and if we know what exact eigenpairs to use, such as in a
sub-cluster approximation. In this case, one should analyze the error
in the approximation, i.e., the error in neglecting the many-body
correlations in $\Heff^\text{c}$. In the perturbative regime, some
results exist \cite{Klein1974}. The author believes, that the
geometric description may facilitate a deeper analysis, and this is an
interesting idea for future work.

\section*{Acknowledgments}

The author wishes to thank Prof.~M.~Hjorth-Jensen, CMA, for helpful
discussions. This work was funded by CMA through the Norwegian
Research Council.

\end{document}